\documentclass[10pt]{iopart}

\usepackage{graphicx}

\date{\today}

\begin{document}
\maketitle

\title{A Bose-Einstein Condensate in an optical lattice}

\author{J~Hecker~Denschlag\footnote[1]{Present address: Institut
f\"ur Experimentalphysik, Universit\"at Innsbruck, Technikerstrasse
25, A-6020 Innsbruck Austria.}, J~E~Simsarian\footnote[2]{Present
address: Bell Laboratories, Lucent Technologies, Holmdel, NJ 07733,
USA.  }, H~H\"affner\dag, C~M$^\mathrm{c}$Kenzie, A~Browaeys,
D~Cho\footnote[3]{Present address: Dept. of Physics, Korea University,
5-1 Ka Anam-dong, Sungbuk-ku, Seoul 136-701, Korea.}, K~Helmerson,
S~L~Rolston, and W~D~Phillips}

\address{National Institute of Standards and
Technology, Gaithersburg, MD 20899, USA.}

\begin{abstract}
We have performed a number of experiments with a
Bose-Einstein condensate (BEC) in a one dimensional optical lattice.
Making use of the small momentum spread of a BEC and standard atom
optics techniques a high level of coherent control over an artificial
solid state system is demonstrated.  We are able to load the BEC into
the lattice ground state with a very high efficiency by adiabatically
turning on the optical lattice.  We coherently transfer population
between lattice states and observe their evolution. Methods are
developed and used to perform band spectroscopy.  We use these
techniques to build a BEC accelerator and a novel, coherent, large-momentum-transfer beamsplitter. 
\end{abstract}


An optical lattice is a practically perfect periodic potential for
atoms, produced by the interference of two or more laser beams. A
Bose-Einstein condensate (BEC) \cite{Anderson1995,Davis1995} is the
ultimate coherent atom source, a collection of atoms, all in the same
state, with an extremely narrow momentum spread. Combining a BEC with
an optical lattice provides an opportunity for exploring a quantum
system analogous to electrons in a solid state crystal but with
unprecedented control over both the lattice and the particles.

Periodic optical potentials have been widely used in atomic physics
(For reviews see \cite{Dowling1996,Grimm2000,Bernet2000}).
Recent experiments and theory have studied BECs in optical lattices
\cite{Anderson1998,Ovchinnikov1999,Kozuma1999,Stenger1999,Deng1999a,Hensinger2001,Morsch2001,Cataliotti2001,Pedri2001,Burger2001,Greiner2001,Orzel2001,Greiner2002}\cite{Berg-Sorensen1998,Jaksch1998,Choi1999,Holthaus2000,Chiofalo2000,Potting2001,Bronski2001}.
Here we present a series of new experiments in which we precisely
manipulate the BEC/lattice system and interpret the results
in terms of a single-particle band structure theory.

We describe 1D lattice experiments with a sodium BEC in a regime where
interactions between atoms are negligible. This paper is organised as
follows: we begin with a description of the experimental arrangement
(Sec.~\ref{expsetup}) and a brief summary of band theory
(Sec.~\ref{theory}). We explore diabatic and adiabatic loading of a
BEC into a lattice in Sec.~\ref{loading}. Using momentum state
analysis we measure coherent superpositions between bands and
determine that we can adiabatically load more than 99\% of the atoms
into a well-defined Bloch state.  In Sec.~\ref{bandtransfer} we
manipulate the lattice potential to transfer population between Bloch
states, mapping the band structure of the lattice eigenstates.  We
follow the work of \cite{Dahan1996,Peik1997,Wilkinson1996} in
Sec.~\ref{blochaccelerator} in the deep lattice regime to develop a
BEC accelerator which imparts many photon recoils of momentum to the
atoms while preserving the coherence of the condensate.  With a little
modification we turn the accelerator into a large momentum transfer
beamsplitter, which is described in Sec.~\ref{beamsplitter}.

\section{Experimental setup}\label{expsetup}

We perform our experiments with a condensate of $3(1) \times 10^6$
sodium atoms~\footnote{All uncertainties reported here are one
standard deviation combined systematic and statistical uncertainties.}
in the $3S_{1/2}$, $F=1$, $m_{F}=-1$ state.  The sample has no
discernible thermal component. The condensate is prepared as described
in \cite{Kozuma1999} and is held in a magnetic time-orbiting potential
(TOP) \cite{Petrich1995,Kozuma1999} trap with trapping frequencies
$\omega_x = \sqrt{2}\omega_y = 2 \omega_z = 2\pi \times $27
Hz. Assuming a scattering length of $a = 2.8$ nm, the calculated
Thomas-Fermi diameters~\cite{Dalfovo1999} are 47, 66, and 94 $\mu$m,
respectively. The size and density of the BEC are such that the
condensate populates on the order of a hundred lattice wells and
interactions between atoms can be ignored on the time-scale of our
experiments.

After production of the condensate the optical lattice is turned
on. It consists of two counter-propagating laser beams along the
$x$-direction whose frequencies and amplitudes can be controlled
independently by acousto-optic modulators (AOMs). The lattice beams
are detuned about $60$ GHz to the blue of the atomic resonance (the
sodium D$_2$ line: $\lambda = 2\pi/k = 589$ nm) so that the
spontaneous emission rate ($\sim 100$ s$^{-1}$) is negligible on the
time scale of our experiments. The
polarization is linear and parallel to the rotation axis of the TOP
trap bias field (the $y$ axis). The condensate is located in the focus
of the beams, which have a $1/e^2$ diameter of about 600 $\mu$m. The
maximum power used in each beam is about 4 mW.

The two counter-propagating laser beams form a standing wave, which
acts on the atoms via the light-shift
\cite{Metcalf1999} to produce a sinusoidal potential:
\begin{equation}
V(x,t) = \frac{V_{0}}{2} \left( 1 +  \cos(2kx + \delta t)
\right). \label{eqpotential}
\end{equation}
Here $\delta$ is the frequency difference between the two laser
beams. This detuning gives the standing wave a velocity of $v =
\delta/2k$.  A frequency difference of $\delta/(2\pi) = 100$ kHz
corresponds to a lattice velocity of $3$ cm$\cdot$s$^{-1}$ (one photon
recoil).

We use time-of-flight analysis \cite{Ovchinnikov1999} to examine the
momentum distributions that result from our experiments. At the
conclusion of the experiment we turn the magnetic trap off
\footnote{It should be noted that, while the magnetic trap is on for
the majority of the experiments, its presence does not
affect our results.} abruptly and the lattice is removed. The details
of the turn-off of the lattice depend on the experiment and will be
described later. Two milliseconds of free-flight allows the momentum
components (which are discrete and separated by 2$\hbar k$ due to the
periodicity of the lattice) to spatially separate. We then measure the
number of atoms in each momentum state by absorption imaging with a
CCD camera.

We adjusted the laser power and detuning such that the height of the
lattice potential $V_{0}$ (measured as described in
Sec. \ref{suddenloading}) was typically 14 photon recoil energies, $E_R
= \hbar^2k^2 / 2M$.  $M$ is the atomic mass of sodium; $E_R/h = 25$
kHz. The total power was constant to within $10\%$.

\section{ Band structure in an optical lattice }\label{theory}

Band structure theory is derived assuming an infinite periodic
potential. Our lattice period ($\approx 0.3\ \mu$m) is small
compared to the Thomas-Fermi diameter (47 $\mu$m) of the condensate in
the lattice direction. Hence, the atomic wavefunction covers more than
a hundred lattice sites and the system can be considered practically
infinite. The corresponding momentum spread of the BEC is very small
($\sim 0.01\hbar k$), so we treat the BEC as a plane wave. Also,
atom-atom interactions are negligible: the time-scale associated with
the peak chemical potential, $\hbar/\mu =
1/(2\pi\times800\;\mathrm{Hz}) = 280$ $\mu$s, is longer than the 20
$\mu$s time scale of our typical experiments. Thus, a single-particle
picture is valid.

The periodicity of the lattice leads to a band structure of the energy
spectrum of the atom-lattice system. The eigenenergies $E_{n,q}$ and
eigenstates $|n,q \rangle$ (Bloch states) of the system in the rest
frame of the lattice are labeled by the quasimomentum $q$ and the
band index $n$. The spatial periodicity of the eigenfunctions results
in a momentum spectrum composed of peaks separated by the momentum
corresponding to the reciprocal lattice vector, $2\hbar k$. The
Bloch states $|n,q \rangle$ can thus be expanded in the discrete plane
wave basis $|\phi_{p}\rangle$ with momenta $p = q + 2m\hbar k$ :
\begin{equation}
|n,q \rangle = \sum_{m  = -\infty}^{\infty} a_{n,q}(m) |\phi_{q +  2 m
\hbar k} \rangle\ . \label{plainwavedecomp}
\end{equation}
In order to calculate the coefficients $a_{n,q}(m)$ and the
eigenenergies $E_{n,q}$ we solve the equation $\hat{H}
|n,q\rangle=E_{n,q}|n,q\rangle$ where the Hamiltonian is $\hat{H} =
\hat{P}^2/2M+ (V_0/2) \cos(2kx)$. 

\begin{figure}
\includegraphics[width=\textwidth]{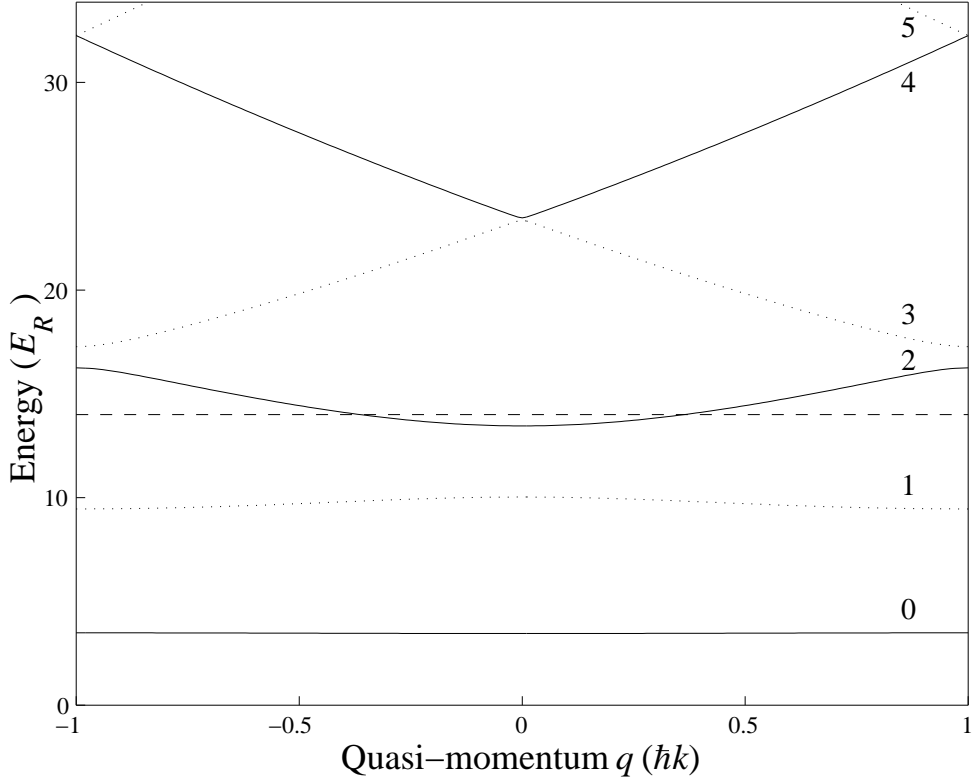}
\caption{ \label{figBigBandDiagram} The first five bands (even bands
are solid, odd bands are dotted) of a 14$E_R$ deep, 1D, sinusoidal
lattice across the first Brillouin zone. The dashed horizontal line
indicates the top of the lattice.}
\end{figure}

Figure \ref{figBigBandDiagram} shows
the eigenenergies versus quasimomentum for the first five bands ($n = 0$
through 4) from $q = -\hbar k$ to $\hbar k$ (the first Brillouin zone) for
$V_0 = 14 E_R $. Even in this relatively shallow potential the
$n=0$ band is quite flat indicating that the tunneling rate is low.

\begin{figure}
\includegraphics[width=\textwidth]{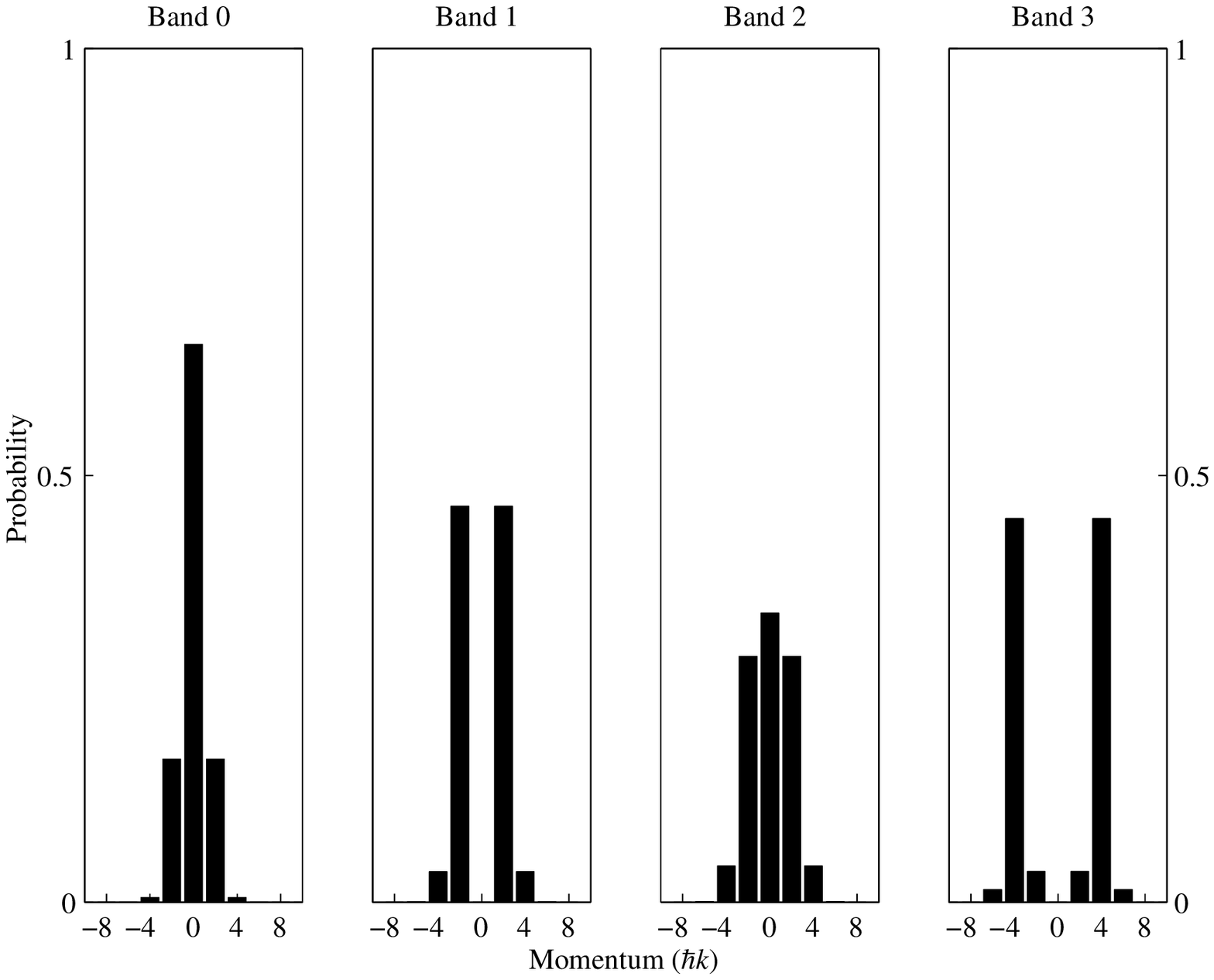}
\caption{ \label{figbanddecomposition} The lattice eigenstates can
be expressed as a superposition of plane waves with fractional
population $|a_{n,q}(m)|^2$. Shown is the plane wave decomposition
of the lattice eigenstates for the four lowest bands for $q = 0$
and lattice height $V_0$ = 14$E_R$. }
\end{figure}

Figure \ref{figbanddecomposition} shows the plane wave decomposition,
$|a_{n,q}(m)|^2$, of the Bloch eigenstates for the bands $n=0$, 1, 2,
and 3 at $q$ = 0 and $V_0 = 14 E_R $. For example, the lattice ground
state $|n=0,q=0\rangle$ consists mainly (65 \%) of a 0 momentum
component and two smaller $\pm 2\hbar k$ momentum components (17 \%
each).

The spatial wave function of a Bloch eigenstate in the lattice
frame, $\psi_{n,q}(x) = \langle x|n,q\rangle$, obeys Bloch's theorem:
\begin{equation}\label{quasimomentum}
\psi_{n,q}(x+\lambda/2) = e^{iq \lambda/2\hbar} \ \psi_{n,q}(x)\ .
\end{equation}
The quasimomentum characterizes the phase difference between
neighbouring lattice sites.

Optical lattices have the advantage that we can easily change their
depth and velocity. In the absence of an external force (or
equivalently an acceleration of the lattice), the quasimomentum is
conserved (even if the depth changes) as long as the lattice
periodicity is kept constant. (Recall that quasimomentum is only
defined in the rest frame of the lattice.)  For an optical lattice
this is obvious since the only mechanism for changing momentum in the
absence of spontaneous emission is redistribution of photons from one
lattice beam to the other. Hence the momentum can only change by
multiples of 2$\hbar k$, which does not change the quasimomentum (see
equation (\ref{quasimomentum})).

\section{Loading the BEC into the lattice}\label{loading}

We now discuss methods for loading the BEC into the lattice with a
selected quasimomentum $q$. We investigate two regimes: the sudden
turn-on and the adiabatic turn-on.

A BEC makes it easy to load a lattice state with a well-defined
quasimomentum. Its small momentum spread (of much less than $\hbar k$)
results in an similar spread of the quasimomentum $q$. For a
condensate stationary relative to the lattice, the phase of the BEC is
uniform across the lattice, hence $q=0$. By contrast, when the
condensate is moving with velocity $v$ relative to the lattice, there
is a linear phase gradient across the condensate and $q=-Mv$.

\subsection{Sudden loading of the lattice}\label{suddenloading}

A BEC, taken to be a plane wave ($|\Psi(t =
0)\rangle=|\phi_q\rangle$), suddenly loaded into a lattice can be
described as a superposition of Bloch states $|n,q\rangle$ :
\begin{equation}
|\Psi(t = 0)\rangle = \sum_{n  = 0}^{\infty}  |n, q \rangle
\langle n,q|\phi_q\rangle\ .
\end{equation}
From equation~(\ref{plainwavedecomp}), $\langle n,q|\phi_q\rangle=
a^*_{n,q}(0)$. Therefore while the BEC wavepacket is held in the lattice
it evolves in time according to
\begin{equation}
 |\Psi(t)\rangle  = \sum_{n = 0}^{\infty} a_{n,q}^*(0) \, \exp\left\{
  -i E_n(q)t/\hbar \right\} |n, q
 \rangle \;.
\end{equation}
We suddenly switch the lattice off after a time $\tau$.
If we project the lattice state onto the plane wave (measurement) basis,
we obtain the coefficients $b_q(m)$ of each $|\phi_{q + 2 m \hbar k}
\rangle$ in the lattice frame:
\begin{equation}
 b_q(m) =  \sum_{n = 0}^{\infty} a^*_{n,q}(0) \ a_{n,q}(m) \
 \exp\left(-i \ \frac{E_n(q)}{\hbar} \ \tau \right).
 \label{eqplainwavecom}
\end{equation}
The interference of the exponential factors produces 
oscillations in the populations of the plane wave components as a
function of $\tau$.

We observe these oscillations as follows: after producing the BEC we
quickly (within 0.5 $\mu$s) turn on the lattice beams and leave them
on for a variable length of time $\tau$. We then shut the lattice off
abruptly (again within 0.5 $\mu$s) and measure, after a time of
free-flight, the relative populations of the momentum components
\footnote{This experimental set up is almost identical to the one
published by Ovchinnikov {\it et al.}  \cite{Ovchinnikov1999}. We
discuss it here again for pedagogical reasons because Ovchinnikov {\it
et al.} use a different model to explain the underlying physics. Also
we have extended that work from measuring at $q=0$ to arbitrary $q$
and our results have significantly improved data quality.}.

\begin{figure}
\includegraphics[width=\textwidth]{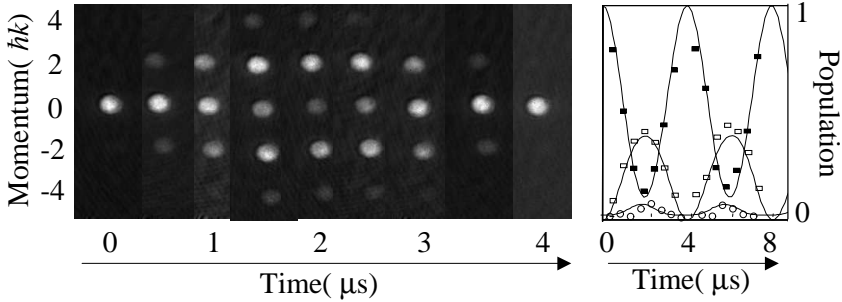} 
\caption{ 
\label{figprojectinlatticeA} Coherent oscillations of the momentum
decomposition of a BEC after
 suddenly switching on the lattice ($q$ = 0, $V_{0} = 14 E_R$).  On
the left are time of flight images showing the plane wave
decomposition of the lattice state evolving as a function of time held
in the lattice. The momentum components have been allowed to spatially
separate. Only a single 4~$\mu$s cycle is shown. The right hand shows
the populations of the 0, +2 and +4$\hbar k$ momentum components
(respectively: the filled squares, open squares and circles) over two
cycles. The lines are a theoretical calculation of the populations in
each component.}
\end{figure}

Figure~\ref{figprojectinlatticeA} shows the time evolution of a BEC
held in the stationary lattice ($q=0$). The populations vary almost
purely sinusoidally, because only bands 0 and 2 are significantly
populated. Band 1 and all odd bands are not populated because the
symmetry of those eigenstates are antisymmetric while the initial BEC
wavefunction is essentially uniform (effectively symmetric) across a
single well.  The observed oscillation period of 250 kHz is the energy
gap between band 0 and 2 for a 14$E_R$ deep lattice. We use this
frequency to calibrate our lattice depth.

\begin{figure}
\includegraphics[width=\textwidth]{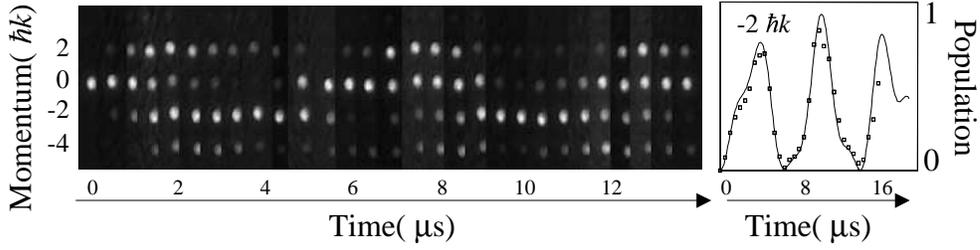} 
\caption{
\label{figprojectinlattice} Coherent oscillations
of a BEC after suddenly switching on the lattice for $q$ = 1$\hbar
k$. This is the same as the experiment shown in figure
\ref{figprojectinlatticeA}, only the quasimomentum has been
changed. This time the oscillations are not sinusoidal. The right hand
side plot shows the evolution of the $-2\hbar k$ component. The
standard 1D band model describes the data well, as shown by the full
line.}
\end{figure}

We can also load the condensate into a lattice moving with a constant
velocity relative to the BEC. In the rest frame of the lattice this
corresponds to loading a specific non-zero $q$. Figure
\ref{figprojectinlattice} shows the results of an experiment, similar
to that shown in figure \ref{figprojectinlatticeA}, for $q=1\hbar k$.
The oscillations are not sinusoidal because more than two bands have
significant population. Odd bands 1 and 3 are now populated in
addition to bands 0 and 2 because the selection rules discussed above
are not valid for $q \neq 0$. The resulting beating signal consists of
a superposition of sinusoids of different frequencies (see equation
(\ref{eqplainwavecom})) corresponding to the energy differences
between pairs of bands. Our observations can be reproduced by the
simple 1D band model, as shown in the right hand side of figure
\ref{figprojectinlattice}.  We have also performed the sudden loading
experiment for other $q$ between 0 and 1$\hbar k$ and the results
agree well with the model\footnote{Note that if we take $q=1\hbar k$
and use a lattice shallow enough that only two bands are populated
then we recreate the Bragg diffraction experiments described in
\cite{Kozuma1999}}.

\begin{figure}
\includegraphics[width=\textwidth]{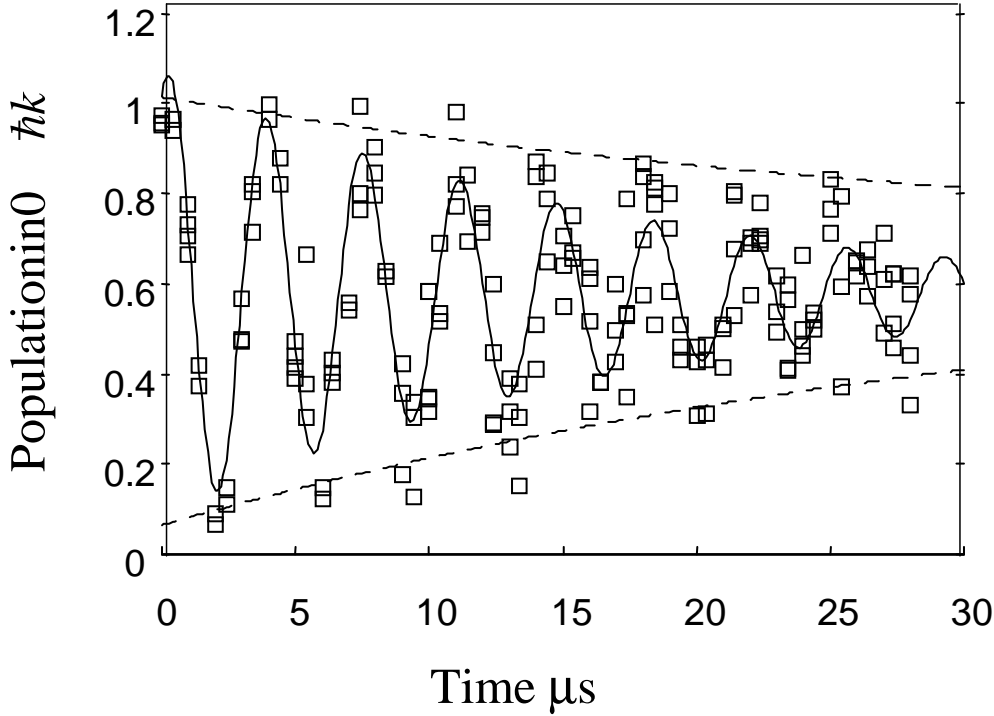}
\caption{ \label{figCohDecay} As the lattice is held on for longer
times the amplitude of the oscillation signal decreases. This is due
to two factors, inhomogeneity of the lattice beams and shot-to-shot
variation in the lattice strength. Both of these effects lower the
oscillation amplitude by averaging several oscillations which are
slightly out of phase with each other. The inhomogeneity of the beams
affects each data point and this decay is the decay of the envelope of
the data points (this is indicated by the dashed lines which have been
drawn to guide the eye). The shot-to-shot intensity fluctuations only
have an effect when the data is averaged or a fit is made. This can be
seen in the solid line which is a fit of a decaying sinusoid to the
data points: the decay of the fit is faster than the decay of the
envelope. Both these effects have to be taken into account when
analysing the data.}
\end{figure}

We have measured the decay time of the beat signal by extending the
lattice interaction time $\tau$. Figure \ref{figCohDecay} shows the
decaying oscillations of the $0\hbar k$ component for $q = 0$. We
fit this data with an exponentially decaying sinusoid with a decay
constant of $17\ \mu$s. Notice that the data scatters widely at long
times. This implies that the decay is due to two mechanisms: the
inhomogeneity ($\approx 10\%$) of our lattice beams over the
condensate and laser power fluctuations from shot to shot (also about
10\%).

\subsection{Adiabatic loading of the lattice}
\label{adiabaticloading}

\begin{figure}
\includegraphics[width=\textwidth]{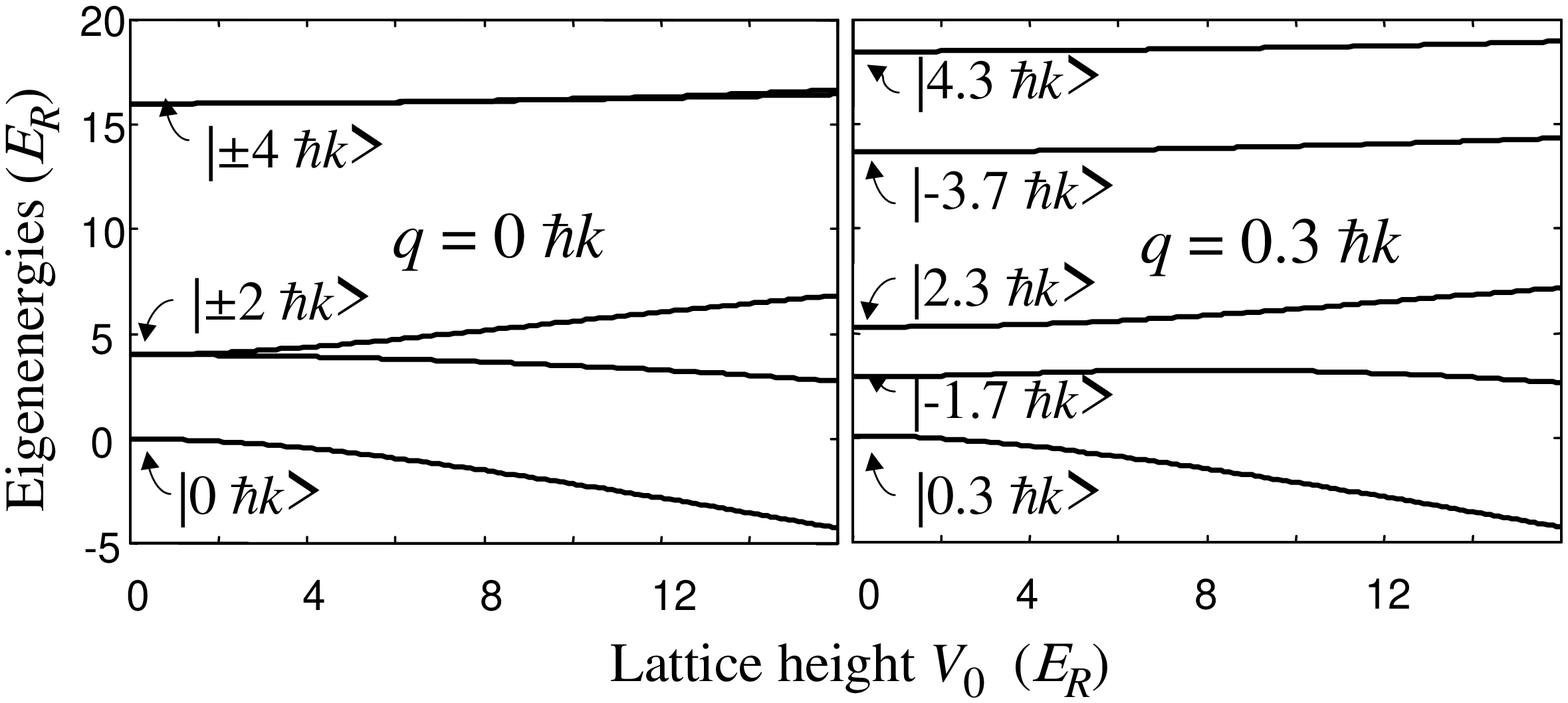}
\caption{ \label{figEVplots} The energy eigenspectrum for two values
of the quasimomentum, $q = 0$ (left) and $q = 0.3\hbar k$ (right),
as a function of the lattice height $V_0$.  At $V_0 = 0$ the lattice
eigenenergies are the free particle kinetic energies, $p^2 / 2M$ and
the eigenstates are the plane waves with momentum $p = 2 m \hbar k +
q$ (in the lattice frame).}
\end{figure}

In contrast to the preceding section, where we populated several bands
by suddenly turning on the lattice, here we transfer the BEC into a
single Bloch state $|n,q\rangle$. To achieve this we ramp up the
lattice intensity adiabatically, i.e., such that \cite{Schiff1968}
\begin{equation}
\label{adiabaticity}
\left|\left\langle i,q \left| \frac{\partial H}{\partial t} \right|
0,q \right\rangle \right| \ll \Delta E^2(q,t)/\hbar.
\end{equation}
where $\Delta E$ is the energy difference between the ground state and
the first excitable state $|i\rangle$. The left hand side is always
less than $\mathrm{d}V_0/\mathrm{d}t$. Therefore the condition in
(\ref{adiabaticity}) can be satisfied at $q=0$ by choosing
$\mathrm{d}V_0/\mathrm{d}t << 16 E_R^2/ \hbar$ since $\Delta E \ge 4E_R$
for any $V_0$ (see figure \ref{figEVplots}). It becomes harder to
maintain adiabaticity as $q$ approaches the Brillouin zone boundary
since there $\Delta E = 0$ when $V_0 = 0$.  The manifestation of the
non-adiabaticity is Bragg scattering. For our experiments, at $q = 0$
with a final lattice depth of 14$E_R$, an adiabatic ramp must be
significantly longer than 5 $\mu$s.

Note that this adiabaticity criterion takes into account only the
single particle motion. We are very non-adiabatic with respect to the
time scale associated with atom-atom interaction energies. Recent
experiments, which observe number squeezing \cite{Orzel2001}
and the Mott transition \cite{Greiner2002}, depend on being adiabatic on
this longer time scale.

\begin{figure}
\includegraphics[width=\textwidth]{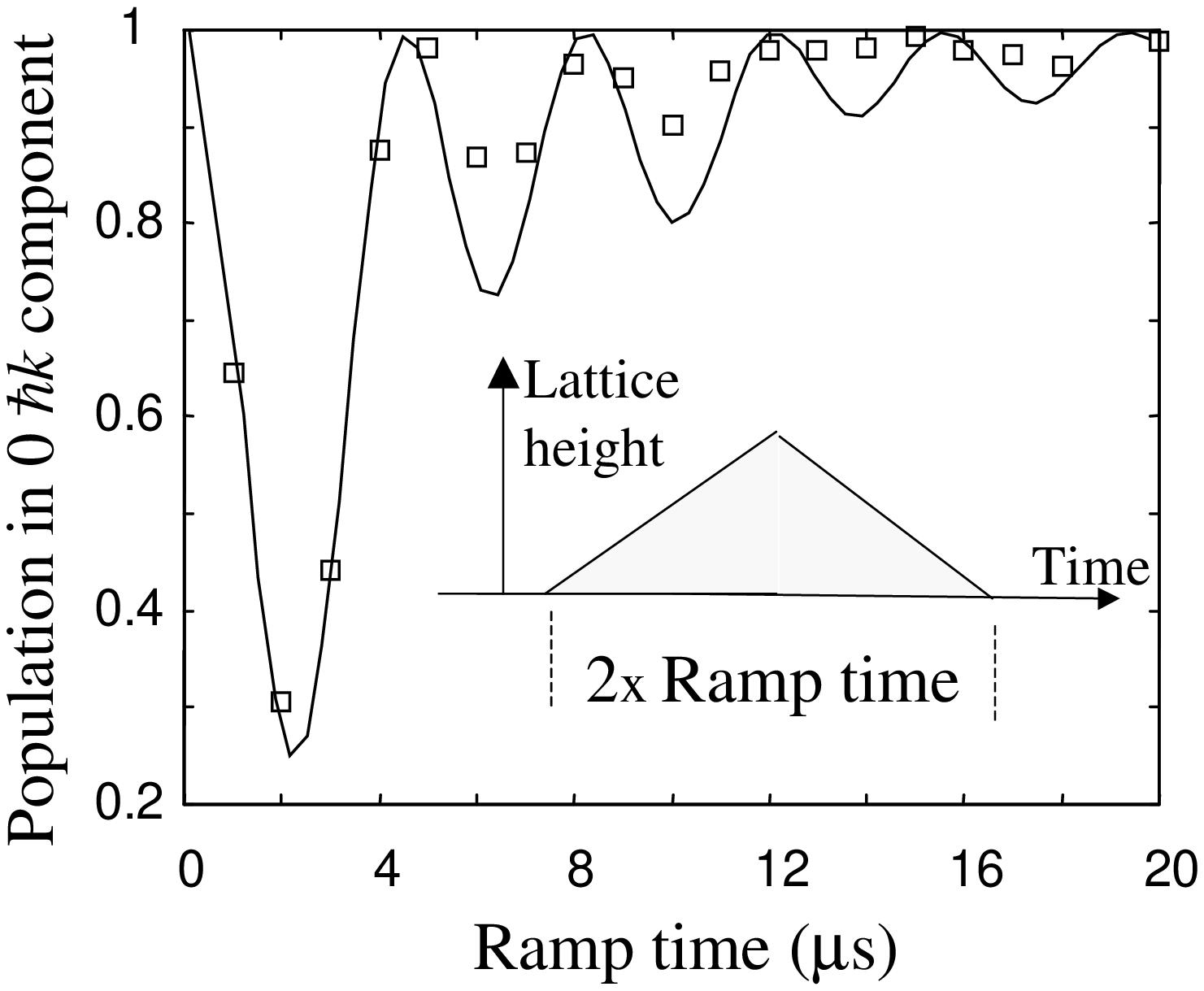}
\caption{\label{doubleramp} Testing the adiabaticity of a linear
intensity ramp with a double ramp. Perfect adiabaticity would leave
all the atoms in the 0 $\hbar k$ state. If the ramp is not fully
adiabatic other momenta are populated. The observed oscillations in
the 0 $\hbar k$ component are due to interference effects similar to
the ones described in section \ref{loading}.  The full line is a
theoretical model obtained through direct integration of the
Hamiltonian. It has been rescaled by about 10\% in the time direction
to match the data.}
\end{figure}

If we can adiabatically ramp up the lattice intensity then we should
also be able to adiabatically ramp down the intensity and return to
the original BEC. This suggests an obvious test for adiabaticity where
we ramp the intensity up and then down over equal time intervals. The
degree of adiabaticity can be determined by measuring the number of
atoms excited into momentum components other than 0 $\hbar k$. The
results of this experiment are shown in figure~\ref{doubleramp}.

This method has two limitations: Firstly, the sensitivity is limited to
the fraction of atoms that can be reliably detected, a few percent in
our case. Secondly, interference between populated bands (as seen in
the previous section) can lead to all the population being in 0~$\hbar
k$ even though the process is far from adiabatic. Because of this, the
signal oscillates and the adiabaticity of a single ramp time can only
be determined by interpolating the envelope of the curve.

\begin{figure}
\includegraphics[width=\textwidth]{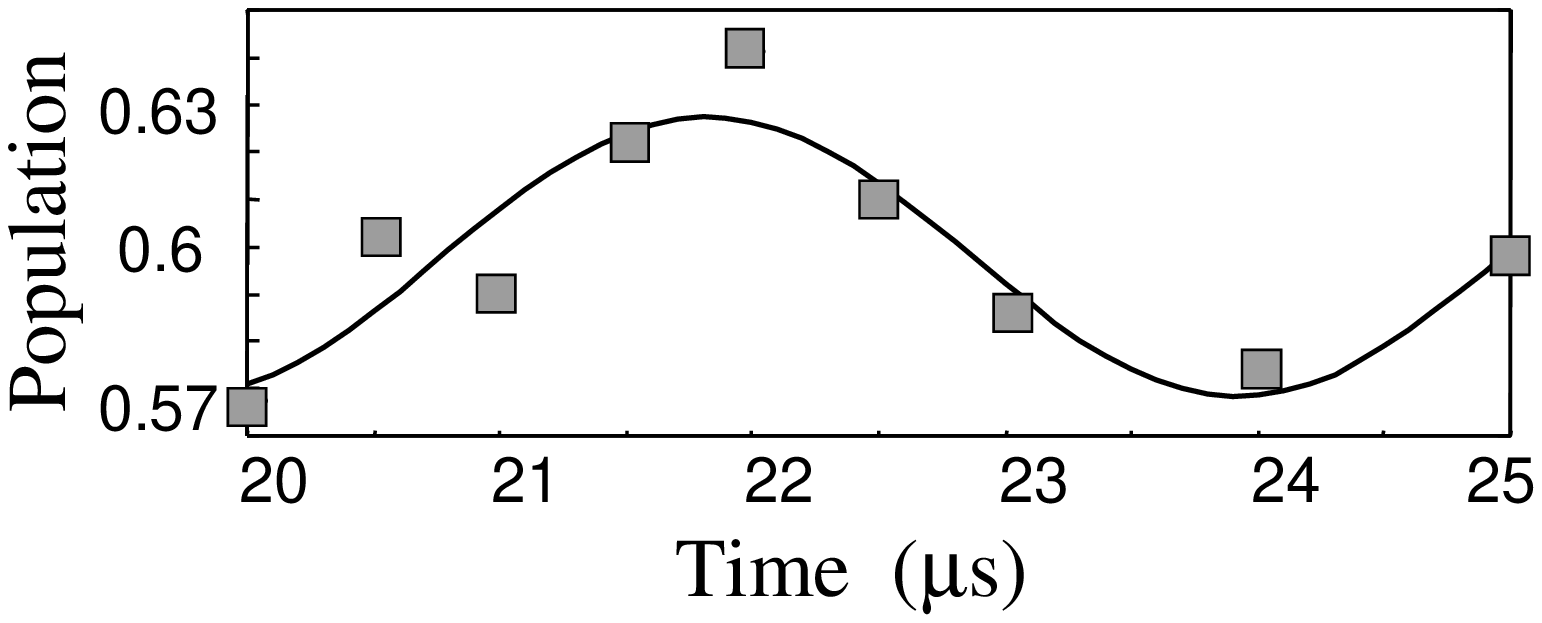} 
\caption{
\label{figstilloscillations} Residual oscillations in the plane
wave decomposition of a lattice state loaded by slowly ramping up the
lattice over $20 \mu$s. The beating is greatly reduced when we raise
the lattice in this manner because the condensate has been loaded
almost entirely into the lattice ground state. This data has been
averaged, but not corrected for loss of amplitude due to inhomogeneity
in the lattice beams or shot-to-shot variations in the lattice height.}
\end{figure}

In order to directly measure the adiabaticity, i.e. the transfer
efficiency into the lattice ground state, we perform experiments
similar to those described in the previous section. Starting from a
condensate at rest we ramp up the intensity of a stationary
lattice\footnote{While we apply a linear voltage ramp, non-linearities in the
response of the AOM smooth the ends ($\approx$10 \% of the ramp time).}. We
hold the BEC in the lattice for a time $\tau$, typically between 0 to
10 $\mu$s, before suddenly switching off the light. We then study the
oscillations in the plane wave decomposition of the lattice
wavefunction as a function of $\tau$.
Figure~\ref{figstilloscillations} shows an example of such
oscillations for the $0\hbar k$ component at $q = 0$ after a ramp time
of 20 $\mu$s. This beating signal has a very small amplitude compared
to the beating signal of figure~\ref{figprojectinlatticeA} for a
lattice that was abruptly switched on.  This indicates that most of
the population is in band 0. If we had populated only the ground
state, there would have been no beating at all. If we ignore the small
oscillations, the measured momentum decomposition (60 \% $0\hbar k$,
20 \% $+2\hbar k$, 20 \% $-2 \hbar k$) is close to the theoretical
decomposition of the lattice ground state (Sec.~\ref{theory}).

From the amplitude of the oscillations we can infer the population in
the lattice ground state. To do this, we assume that only band 0 and 2
are populated (since bands 1 and 3 cannot be excited at $q = 0$). Let
$p$ be the fraction of the population in band 0 after loading into the
lattice. Then the population in band 2 is $1-p$.  For a lattice with a
height of $14 E_R$ band 0 has a calculated $0\hbar k$ momentum
component of 65 \% whereas band 2 has one of 34 \% (see figure
\ref{figbanddecomposition}). The fraction of population in the $0\hbar
k$ momentum component after evolving for a time $\tau$ is 
\begin{equation}
P(\tau) = \left| \sqrt{0.65p} + 
e^{i(\omega\tau + \theta)}\sqrt{0.34(1-p)}\right|^2\ \,
\end{equation}
where $\omega$ is the frequency difference between bands 0 and 2 and
$\theta$ is a phase that is dependent on the details of the ramp. The
amplitude of the beating is
\begin{equation}
\label{deltap}
\Delta P =  2\sqrt{0.65p \times 0.34(1-p)}.
\end{equation}
Equation (\ref{deltap}) does not take into account the decay of the
beating amplitude due to dephasing. Before calculating $p$ from the
experimentally measured value for $\Delta P$, we must first correct
for the dephasing. We cannot directly use the decay rate obtained from
figure \ref{figCohDecay} since, in this case, we use an intensity ramp
rather than a constant intensity. To find the correct decay rate a
simulation of the sudden turn on was performed with the effects of
shot-to-shot noise and beam inhomogeneity included. The simulation was
a direct solution of Schr\"odinger's equation with no mean-field
effects. The parameters of the dephasing effects were then adjusted to
match the observed decay and were then applied to a similar simulation
which included the intensity ramp.

\begin{figure}
\includegraphics[width=\textwidth]{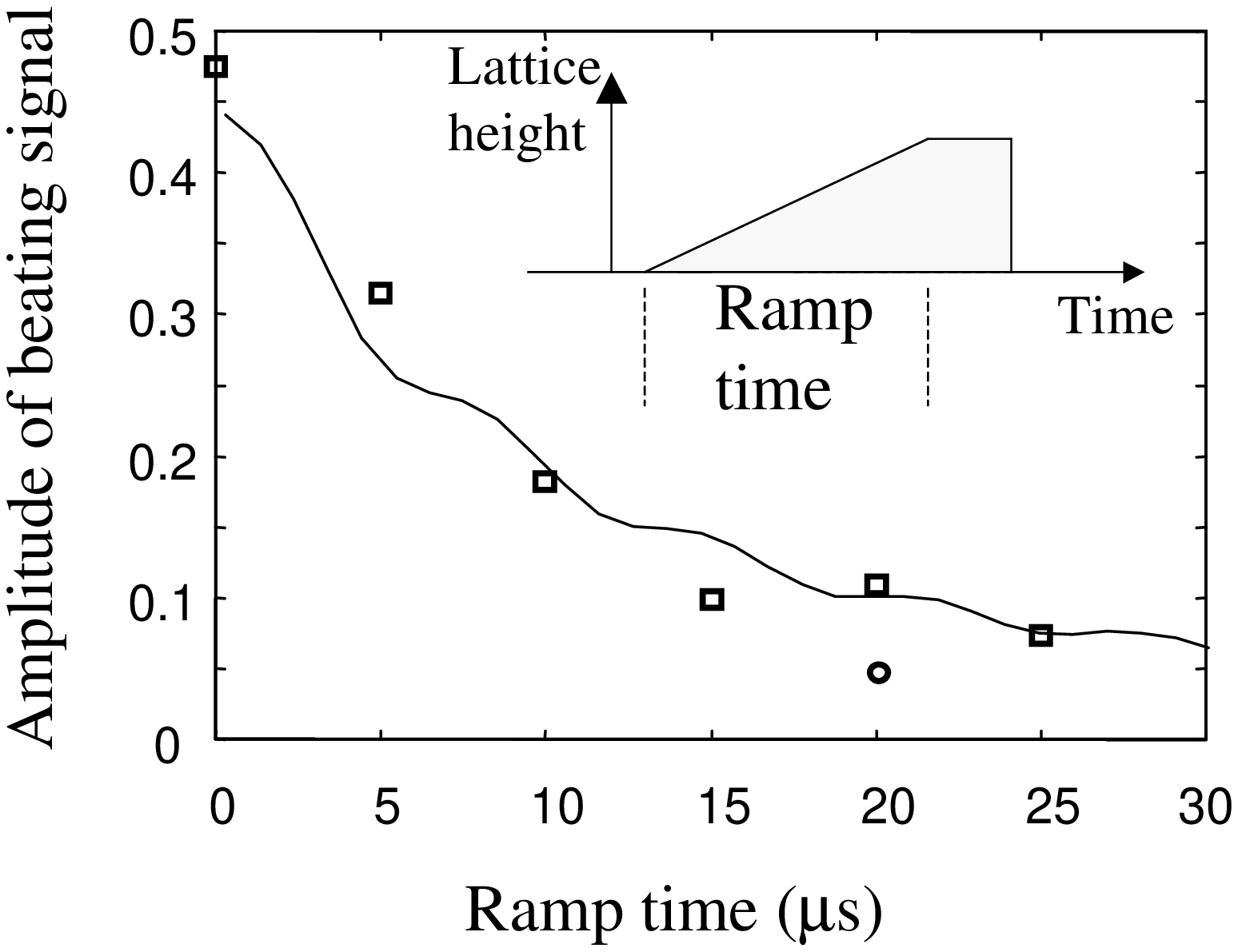}
\caption{\label{figsinglramp} The decrease of the beating signal with
increasing ramp time for a final lattice height of 14$E_R$. The data
points are corrected to account for loss of coherence. The full line
is from a band-theory calculation for a lattice with $V_0 = 14 E_R$.
The insert sketches how the lattice height is ramped
during the experiment. The open circle 
corresponds to the data in figure~\ref{figstilloscillations}.}
\end{figure}

Figure~\ref{figsinglramp} shows how the oscillation amplitude
decreases as a function of the lattice ramp time. The data are
corrected for decay as described above. The largest correction is a
factor of 1.8 increase in the oscillation amplitude. The data agree
well with a calculation (full line) where the single-particle
Hamiltonian corresponding to loading a $V_0 = 14E_R$ lattice is
numerically solved. The lowest amplitude shown in figure
\ref{figsinglramp} (the open circle, this data is shown in
figure~\ref{figstilloscillations}) corresponds to a ramp time of 20
$\mu$s and a groundstate population of 99.7(1)\% (the uncorrected
result is 99.9\%).

Finally we point out that a linear ramp is not the most
efficient way to populate the ground state. Numerical simulations
\footnote{Carl Williams, private communication.} show that our actual
ramp (smoothed by the AOM response) is more adiabatic than a linear
ramp. These simulations agree well with our observed
adiabaticity. Other ramp shapes could be made even more
adiabatic. Since the band gap increases with the lattice intensity,
the ramp can be accelerated as the intensity increases.

\setcounter{footnote}{0} 

As an alternative we find that we can dramatically improve the loading
of the ground state by using interference effects. The single linear
ramp can for example be broken up into two linear ramps with different
slopes. Both ramps coherently drive population between the ground and
excited states. The second ramp can then be used to almost entirely cancel
the effect of the first ramp\footnote{Our numerical calculations show
that using two optimized linear ramps one can achieve a ground state
population greater than 99.98 \% for a total ramp time of 20 $\mu$s
and $V_0 = 14 E_R$.  This value for the ground state population is
even stable to lattice intensity fluctuations of 10\%.}. Details of
these studies will be subject of future work.

\section{Coherent Transfer between Lattice Bands and
 Band-spectroscopy}\label{bandtransfer}

Starting from a condensate adiabatically loaded (20 $\mu$s ramp) into
the ground state of the lattice, we will now study coherent population
transfer to higher bands.

\subsection{Transfer between bands 0 and 1}\label{transferband01}

\begin{figure}
\includegraphics[width=\textwidth]{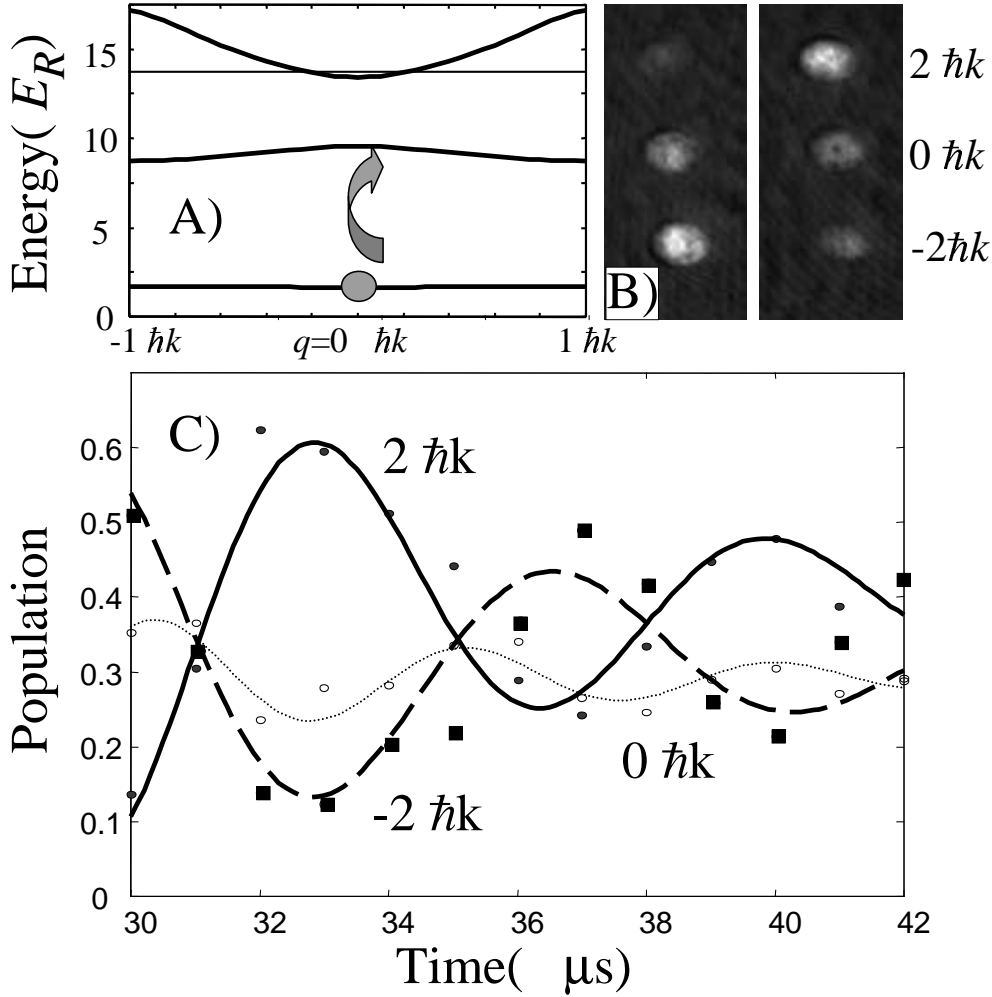}
\caption{ Starting from \label{figcohereTransfer} the lattice
ground state ($q = 0$, $n$ = 0) population can be coherently transferred
to the band 1 (at $q = 0$) by shaking the lattice for 30 $\mu$s. This is
illustrated in A. In B and C the oscillations
in the plane wave decomposition of the lattice state are shown after
transferring some population to band 1.
Oscillations are strongest between the $\pm 2 \hbar k$ components.
Fitting exponentially decaying sinusoids (C) to the
oscillating momentum components yields an oscillation frequency of
140 kHz for the $\pm 2 \hbar k$ components.}
\end{figure}

To transfer population from band 0 to 1 at $q = 0$ (see figure
\ref{figcohereTransfer}) we modulate the phase of one of the lattice
beams and thus the position of the wells, i.e. shaking the
lattice. The modulation is performed using an AOM and has an
amplitude of $\approx \pi/6$. We apply the modulation for four cycles
over approximately 30 $\mu$s.

Shaking the lattice is an odd-parity excitation and therefore
efficiently couples bands of opposite parity, such as bands 0 and
1. The phase modulation can also be viewed as putting sidebands on the
lattice beam. In this picture the transition between bands is a Raman
transition and the ability to make odd parity excitations is related to
the phase difference of $\pi$ between the sidebands.

A convenient way to analyze the population of excited bands is to
adiabatically ramp down the lattice after the shaking.  Population in
band 1 at $q=0$ is all transferred, after the ramp-down, into the $\pm
2\hbar k$ components. The frequency of the 0 to 1 transition can
then be found by maximising the population in these momentum
components.

Once the resonance is found we can examine the population transfer via
interference experiments. If we switch off the lattice suddenly we
observe oscillations as a function of the holding time (see
figure~\ref{figcohereTransfer} B and C). Since bands 0 and 1 have
opposite parity the oscillations will tend to be between momentum
components with opposite momenta (in contrast to the beating of bands
0 and 2 illustrated in figure \ref{figprojectinlatticeA}). Fitting an
exponentially decaying sinusoid to the oscillating momentum components
yields an oscillation frequency of 140 kHz for the $\pm 2\hbar k$
components consistent with the calculated frequency of 147 kHz for the
11$E_R$ lattice used for this experiment. The oscillations show that
we have coherently populated band 1. The small oscillation of the 0
$\hbar k$ momentum peak is due to population in band 2 (the
oscillation frequency corresponds to the energy gap between bands 0
and 2) that has been excited by second-order processes.

\subsection{Transfer between bands 0 and 2}
\label{transfer1and3}

We can transfer population between bands 0 and 2, without exciting
band 1, by amplitude-modulation of the lattice, an even-parity
excitation. This can be viewed as a Raman process as well; in this case
the sidebands are in phase.

\begin{figure}
\includegraphics[width=\textwidth]{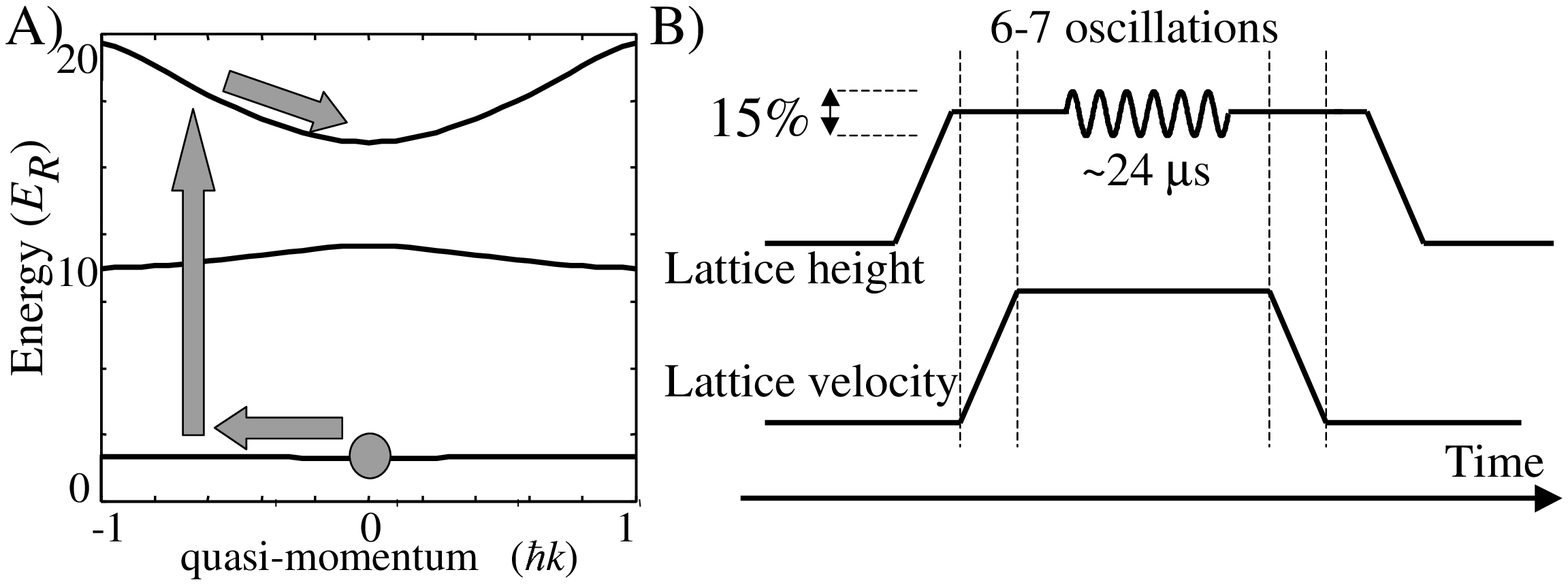}
\caption{
\label{figbandspectroScheme} The scheme for transferring population
into band 2 at arbitrary $q$. We start from the ground state at a
quasimomentum of 0. We then accelerate the lattice to change the
quasimomentum from 0 to the desired value of $q$.  Population is
transferred into band 2 by modulating the lattice height over a 20 to
30~$\mu$s period.  In order to measure the population in the upper
band the acceleration is reversed to bring the quasimomentum back to 0
and the lattice is adiabatically released to map the bands into
individual momentum components.}
\end{figure}

Although band 1 is relatively flat, band 2 has significant
curvature (see Figs.~\ref{figBigBandDiagram} and
\ref{figbandspectroScheme} A) which can be examined by varying $q$ and
measuring the resonance frequency. The resonance is found by varying
the modulation frequency to maximise the loss from band 0.

To produce a non-zero $q$ within the first Brillouin zone we start
from an adiabatically loaded lattice ground state ($q = 0$) and then
suddenly apply a constant acceleration $a = 300$~m$\cdot$s$^{-2}$ to
the lattice. For $|Ma\langle 1,0 | x | 0,0 \rangle| \ll \Delta E$
(corresponding to $a<3.9\times10^4\;m\cdot s^{-2}$ in our case)
this sudden change does not significantly excite higher bands. In the
lattice frame the inertial force changes $q$ according to
$\mathrm{d}q/\mathrm{d}t = Ma$.  We adjust the duration of the
acceleration to obtain the desired $q$. (We could have adiabatically
loaded the condensate directly into a moving lattice to produce a
non-zero $q$, but achieving adiabaticity is difficult near $q = 1\hbar
k$.)

\begin{figure}
\includegraphics[width=\textwidth]{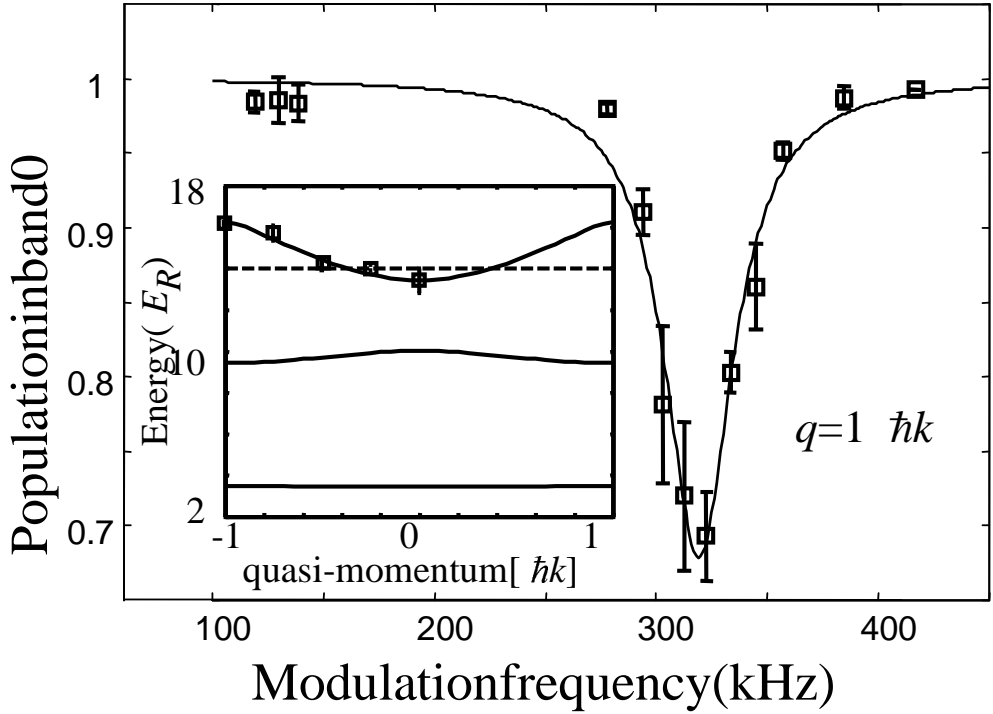} 
\caption{ 
\label{figBandSpectResult} Experimental data showing population
transfer from band 0 to band 2 at $q = 1\hbar k$ as a function of excitation
frequency. The scheme used is outlined in
figure~\ref{figbandspectroScheme}. The fit is a Lorentzian. The
resonance frequency corresponds to the band gap between band 0 and 2
at $q = 1\hbar k$. Similar measurements at $q$ = 0, 0.25, 0.5 and 0.75$\hbar
k$ allow us to map out the curved band structure (see insert).}
\end{figure}

We suddenly stop accelerating, holding $q$ constant.  The lattice
height is then modulated by 15\%\ (controlled with an AOM) for 6 to 7
oscillations. Using this relatively short and strong agitation to
transfer the population broadens the transition and makes it easier to
find the resonance line. In order to measure how much population has
been transferred to band 2, the lattice is decelerated to zero
velocity ($q = 0 $) and then adiabatically ramped down to map the
lattice bands into plane waves (see figure \ref{figEVplots}). Figure
\ref{figBandSpectResult} shows the resonance curve for $q = 1\hbar
k$. The resonance frequency at about 320 kHz equals the expected band
gap between band 0 and 2 at $q = 1\hbar k$. The width of the
excitation is determined by the bandwidth of the excitation ($\approx
37$ kHz). In the insert we plot the measured resonance frequencies
for the quasimomenta $q$ = 0, 0.25, 0.5, 0.75 and 1$\hbar k$ together with
the calculated band structure for $V_0 = 14E_R$. The
data agrees well with the band model.

\section{An accelerator for a BEC }
\label{blochaccelerator}

An accelerated lattice can be used to reach velocities corresponding
to momenta well beyond the first Brillouin zone (Similar work with
thermal atoms has been described in \cite{Peik1997}). Using the method
described above we adiabatically load the BEC into a 14$E_R$ deep
lattice with a 25 $\mu$s ramp. We then accelerate the lattice at 1600
m$\cdot$s$^{-2}$. This sweeps the quasimomentum through a Brillouin zone
every 35 $\mu$s. 

\begin{figure}
\includegraphics[width=\textwidth]{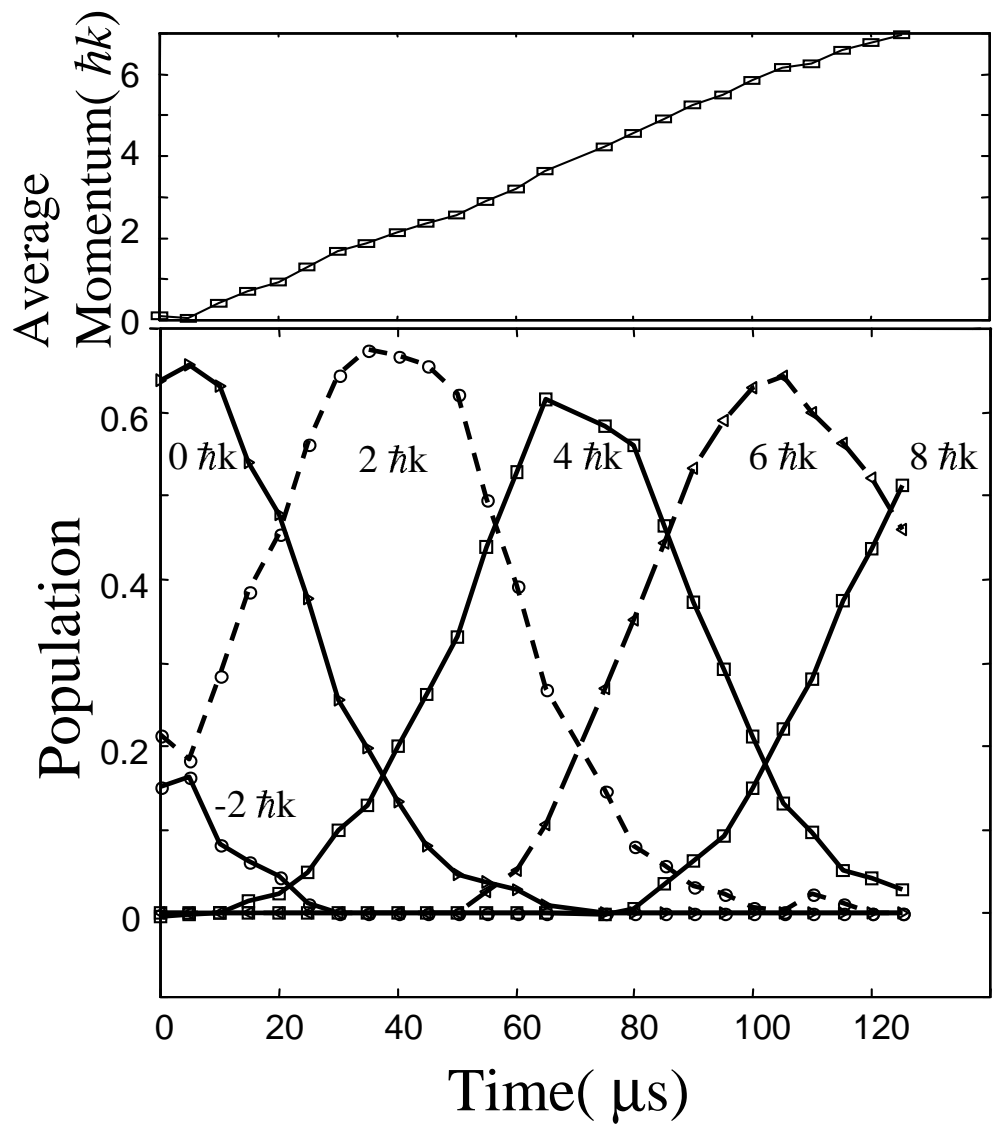}
\caption{ \label{figblochoscillation} The momentum decomposition and
mean momentum of a condensate in the lowest band after a period of
constant acceleration. The acceleration sweeps the atoms through a
Brillouin zone in 35$\mu$s. The reference frame is the stationary
laboratory.}
\end{figure}

Figure \ref{figblochoscillation} shows the change of the mean momentum
and the momentum distribution of the accelerated BEC in the laboratory
frame measured by suddenly switching off the lattice after a given
period of acceleration. We see a linear increase of the mean momentum
(corresponding to the lattice velocity). Note that despite the
continuous change of the mean momentum, individual atomic momenta
(measured in the lab frame) are always multiples of $2\hbar k$. After
passing through each Brillouin zone momentum decompositions are
similar and recurring, shifted only by multiples of $2\hbar k$. This
recurrence is due to the recurrence of $q$ as it scans the Brillouin
zone from $-\hbar k$ to $\hbar k$ and then undergoes Bragg scattering
back to $-\hbar k$.

Despite the similarity of our experiments to those which exhibit Bloch
oscillations in the group velocity
\cite{Dahan1996,Wilkinson1996,Morsch2001} the oscillations in our case
are too small to be observed.  The group velocity of the atoms in the
lattice frame (given by $v_g = \mathrm{d}E/\mathrm{d}q$) is, in our
case, at most 7.3$\times10^{-4}$ m$\cdot$s$^{-1}$, which is negligible
compared to the lattice velocity. (The maximum lattice velocity shown
is 2.1$\times10^{-1}$ m$\cdot$s$^{-1}$.) With
increasing potential the Bloch oscillations in the group velocity are
suppressed and we are left only with the oscillations in the momentum
decomposition.

\begin{figure}
\includegraphics[width=\textwidth]{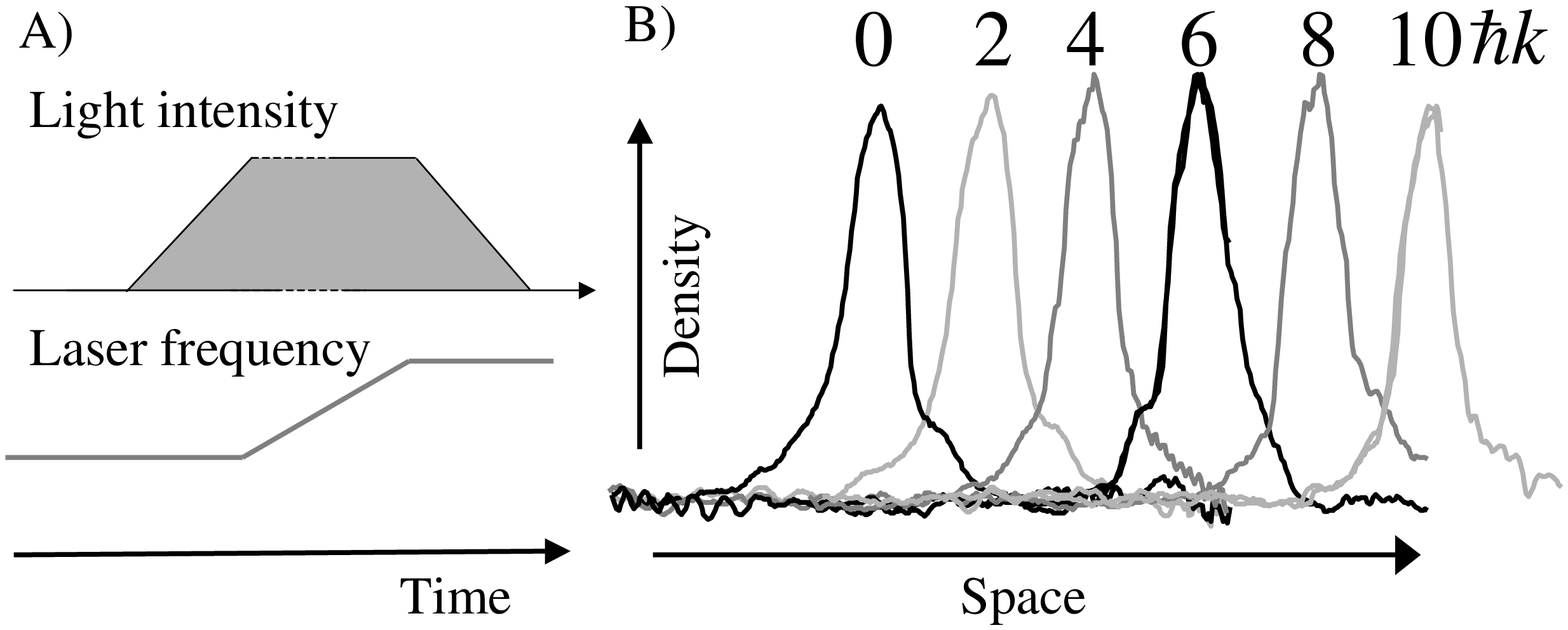}
\caption{ \label{figblochacceleratorB} Accelerator for cold atoms.
First the BEC is adiabatically loaded into the lattice ground state
($q = $0) and the lattice is subsequently accelerated (A). After
sweeping through an integer number $m$ of Brillouin zones, the lattice
is ramped down adiabatically, which results in a single BEC wavepacket
moving at $2m \hbar k$. (B) shows accelerated BEC wavepackets with
final momenta of up to 10$\hbar k$.}
\end{figure}

To obtain an accelerated BEC with a single momentum we turn the
lattice off adiabatically. The BEC has a final momentum of $2 m\hbar
k$ where $m$ is the number of Brillouin zone boundaries crossed during
the acceleration (i.e. the final lattice velocity is between
$(2m-1)\hbar k/M$ and $(2m+1)\hbar k/M$.). We have
accelerated a BEC up to momenta of 10$\hbar k$ (see figure
\ref{figblochacceleratorB} B). The final velocity is quite insensitive
to power and frequency fluctuations in the lattice beams and to the
specifics of the lattice acceleration. The limitations on our final
velocity are purely technical: the range of the frequency sweep and
the field of view of our camera.

While there is no apparent limit on the final velocity there are
limits on the acceleration. In our case the sudden initiation of the
acceleration presents a limit as described above. If we adiabatically
applied the acceleration to circumvent this limit there will always be
the ultimate limit of ``spilling'': when the acceleration is so large
that there are no bound states.

\section{A large-momentum-transfer beamsplitter}
\label{beamsplitter} 

We now describe how we use the accelerator in order to build a
coherent, large-momentum-transfer beamsplitter (LMT-beamsplitter).
This is useful for applications that require atoms with large,
coherent separations (in either position or momentum). A
LMT-beamsplitter can be built using high-order Bragg diffraction
\cite{Kozuma1999}, but is limited to relatively low momenta because
the transition probability decreases rapidly with increasing
order. Our approach is not limited in this way. It is similar in
spirit to coherent beamsplitters based on adiabatic transfer
\cite{Lawall1994,Goldner1994,Weitz1994}.

\begin{figure}
\includegraphics[width=\textwidth]{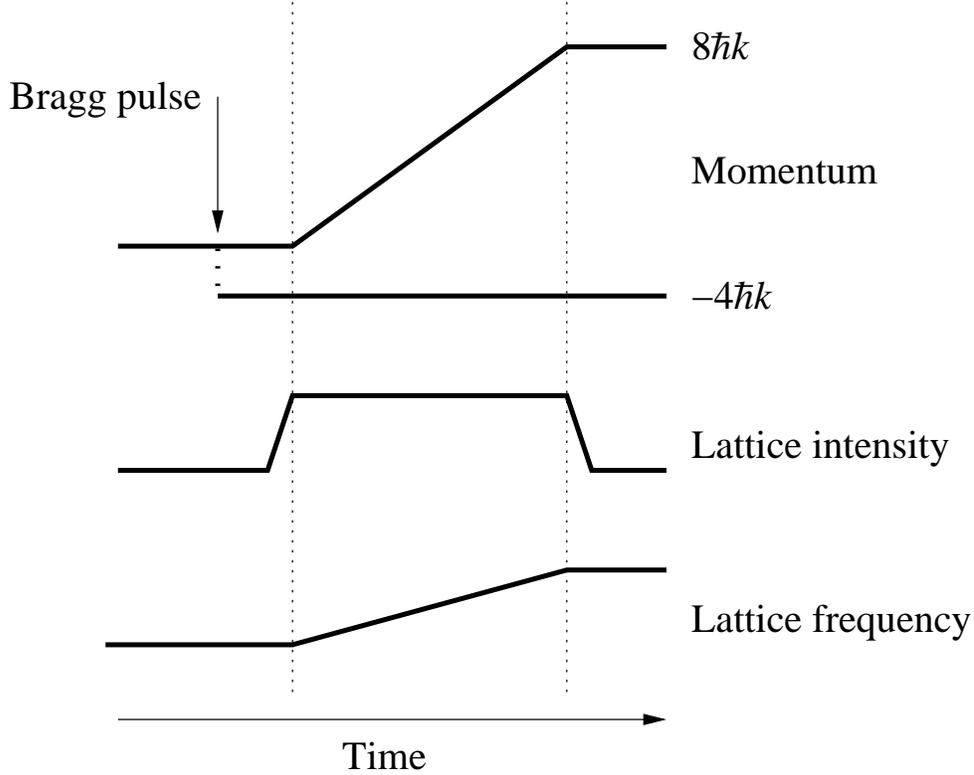}
\caption{ \label{figCohBeamsplitter} A coherent, large-momentum,
transfer beamsplitter.  After splitting the condensate with a
second-order Bragg pulse, a lattice moving at $0.3\hbar k/M$ is
adiabatically turned on over 20 $\mu$s, such that the $0\hbar k$ and
$-4\hbar k$ components are transferred into bands 0 and 4,
respectively. The lattice is then accelerated over 200 $\mu$s. The
component in band 0 accelerates with the lattice, while the component
in band 2 stays at rest. After sweeping through $m$ Brillouin zone
boundaries, the lattice is adiabatically turned off, resulting in two
wavepackets, one with momentum -4$\hbar k$ and one with a momentum of
$2m \hbar k$. In our experiments we swept through up to 4 Brillouin
zone boundaries in about 200 $\mu$s.}
\end{figure}

Our scheme is outlined in figure~\ref{figCohBeamsplitter}: We begin
with a second-order Bragg pulse \cite{Martin1988,Kozuma1999} that produces
a coherent superposition of the two momentum states 0 and -4$\hbar
k$. We use a $8E_R$ high lattice pulsed on for 6.4 $\mu$s at a
velocity of $-2\hbar k/M$ to perform the splitting. We then
adiabatically turn on a 14$E_R$ deep lattice at a velocity of
$0.3\hbar k/M$. The lattice is accelerated to a velocity of
approximately $2m\hbar k/M$ carrying the 0$\hbar k$ momentum component
along with it. By contrast, the -4$\hbar k$ momentum component remains
stationary.

The selective acceleration is achieved by choosing a lattice velocity
such that the atoms with momentum 0$\hbar k$ in the lab frame are in
band 0, while the atoms with -4$\hbar k$ are in band 4. Both have a
quasimomentum of $-0.3 \hbar k$. We chose this $q$ to ensure there are
no degeneracies between bands at $V_0 = 0$ (see right hand side of
figure \ref{figEVplots}).

As we accelerate the lattice, the atoms in band 0 are accelerated with
the lattice, just as described in the previous section. Crossing the
Brillouin zone boundary does not change the band. By contrast the
atoms in band 4 respond as nearly free particles (since band 4 is
unbound): when $q$ reaches the avoided crossing between bands 4 and 5
at the zone boundary (see figure \ref{figBigBandDiagram}) the atoms
pass through to band 5 diabatically. At subsequent level crossings
(such as for bands 5 and 6 at $q=0$) this process continues. In the
lab frame this corresponds to the atoms being unaccelerated.

We can apply the Landau-Zener criterion for diabatic passage through
the avoided crossings. The probability for a passage between bands
$(n-1)$ and $n$ is \cite{Zener1932} :
\begin{equation}
\label{eqzener}
 P_{\mathrm{LZ}} = \exp{\left\{- \pi  \frac{\omega^2}{4\, n\,k\,
 a} \right\}},
\end{equation}
where $\omega$ is the band gap frequency. This expression is derived
in the limit where the anti-crossings are small. For the passage from
bands 4 to 5 the gap is 2.2 kHz, giving a diabatic probability of
99.8\% for our acceleration of $a=1200$~m$\cdot$s$^{-2}$. (If we had
used a first-order Bragg pulse to initially split the condensate into
0 and 2$\hbar k$ components, then only 72\% of the population would
have passed through the first crossing.)

We end the experiment by adiabatically ramping down the intensity of
the moving lattice.  We get two wave packets, one at $2m\hbar k$ (recall
that $m$ is the number of Brillouin zone boundaries crossed), the
other at rest. We have demonstrated splittings of up to 12$\hbar k$.

The LMT-beamsplitter can be used to construct an interferometer.  We
have performed a proof-of-principle experiment in which we have built
a Mach-Zehnder interferometer similar to that described in
\cite{Denschlag1999,Torii2000}. In our case the initial Bragg pulse
beamsplitter was replaced with the LMT-beamsplitter and the final
Bragg pulse beamsplitter with a time-reversed version of the
LMT-beamsplitter. In addition the Bragg $\pi$-pulse of the
conventional Mach-Zehnder interferometer is replaced with a three
stage LMT-$\pi$-pulse consisting of a deceleration stage, a
conventional Bragg $\pi$-pulse and a final acceleration stage. The
resulting interference pattern had a contrast of 50\%, which was
limited by transverse motion of the BEC during the experiment.

\section{Conclusions}
In conclusion, we have performed a number of experiments with a BEC in
static, moving, and accelerating 1D optical lattices and have
interpreted the results using band structure. We have studied
adiabatic loading into the ground state and coherent transfer to other
lattice states by modulating the lattice. These techniques have been
used to perform band spectroscopy. We have also built an coherent
accelerator and a multiphoton beamsplitter.

During preparation of this paper we learned of several other optical
lattice experiments with a BEC
\cite{Morsch2001,Cataliotti2001,Pedri2001,Burger2001,Greiner2001,Greiner2002}.

\ack
We thank Keith Burnett, Mark Edwards, Lu Deng and Ed Hagley for
fruitful discussions. This work was supported in part by the US
Office of Naval Research, NASA and ARDA/NSA. J.~H.~D. and H.~H. acknowledge
support from the Alexander von Humboldt foundation. A.~B.
acknowledge partial support from DGA.

\section*{References}

\bibliographystyle{unsrt}
\bibliography{lattices}

\end{document}